\documentclass[11pt]{article}
\usepackage[english]{babel}
\usepackage{amssymb}
\def\ba{\begin{eqnarray}}
\def\ea{\end{eqnarray}}
\def\nn{\nonumber}
\def\fy{\varphi}
\def\fyd{\varphi^{\dagger}}
\def\gt{\delta_{g}}
\begin{document}
\title{{Generalized Gauge Transformations and Regularized
$\lambda\varphi^{4}$-type  Abelian Vertices}\author{Winder A.
Moura-Melo$^{\mbox{a,b}}$ \thanks{Email: winder@stout.ufla.br, winder@cbpf.br}
\hskip .1cm and J.A. Helay\"el-Neto$^{\mbox{b,c}}$ \thanks{Email:
helayel@cbpf.br.} \\ \\$^{\mbox{a}}$\hspace{.1cm}Deptartamento de Ci\^encias
Exatas, Universidade Federal de Lavras\\ Caixa Postal 37, 37200-000,
Lavras, MG, Brasil\\ $^{\mbox{b}}$\hspace{.1cm}Centro Brasileiro de
Pesquisas F\'{\i}sicas\\ Rua Xavier Sigaud 150 - Urca, 22290-180 - Rio de
Janeiro, RJ - Brasil. \\$^{\mbox{c}}$\hspace{.1cm} Grupo de F\'{\i}sica
Te\'orica, Universidade Cat\'olica de Petr\'opolis\\Av. Bar\~ao do
Amazonas 124, 25685-070, Petr\'opolis, RJ, Brasil.}}
\date{} 
\maketitle 
\begin{abstract} Abelian Lagrangians containing
$\lambda\varphi^{4}$-type vertices are regularized by means of a suitable
point-splitting scheme combined with generalized gauge transformations.\ The
calculation is developped in details for a general Lagrangian whose fields
(gauge and matter ones) satisfy usual conditions. We illustrate our
results by considering some special cases, such as the
$(\overline{\psi}\psi)^{2}$ and the Avdeev-Chizhov models. Possible
application of our results to the
Abelian Higgs model, whenever spontaneous symmetry breaking is considered, is
also discussed.\ We also pay attention to a number of features of the
point-split action such as the regularity and non-locality of its new
``interacting terms''. \end{abstract} 
\section{Introduction} In the quantum field-theoretical framework of modern
Physics, products of fields at the same space and/or time point are not
well-defined since these fields are taken as operator-valued distributions.
As a consequence of such ill-defined products, we are led to divergent
results when calculating relevant quantities, such as, physical masses and
coupling constants. Physically speaking, such divergences arise
because we describe elementary particles as if they were point-like entities
and, consequently, carrying infinite density of mass,
charge, and so forth.\\ \\
Even though there are several regularization methods to deal with such problems,
those based on point-splitting may offer some advantages respect to others when
performed in a suitable way.\ Essentially, the procedure works by taking the
field products initially at the same point, and later on at different points,
by {\em splitting} them.\ The result is such that the new Lagrangian contains
only regularized interaction terms.\ Already in 1934, Dirac employed
such an idea in order to split products of quantities at the same points
which appeared in density matrices of electronic and positronic  physical
distributions (see Ref.\cite{Dirac}, for details) .\\ \\
More recently, several results have been obtained by means of this method for
both Abelian (QED) and non-Abelian (Standard Model) cases.\ For example, the
values of some important physical parameters such as the top quark and Higgs
scalar masses, have been got free from divergences and were shown to be in good
agreement with other schemes.\ This procedure
was also shown to respect the gauge invariance of the theories (for details
see the papers listed in Refs. \cite{OW,OW2}).\\ \\
Nevertheless, these works did not pay enough attention to the explicit
construction and form of the new {\em point-split} gauge
transformations.\ Such an issue was the subject of a more recent
paper, Ref.\cite{GNW}, where the Abelian infinitesimal
form of these new transformations (the so-called {\em generalized gauge
transformations}, denoted by ggt's) was proven to exist to all orders in the
gauge coupling constant.\ The explicit forms of such ggt's  as well as of a
generalized QED-Lagrangian were presented up to fourth order.\ This new
Lagrangian, obtained from the original QED action, was shown to be regularized,
i.e., its interaction terms (including some new ones which appear from the
splitting) presented no product of fields at the same point. On the other
hand, those new terms also displayed non-locality property.\ As expected, as
we set the point-splitting parameter to zero, we recover the original
results.\\ \\ 
Although the ggt's have been built up for QED, we do not see any
restriction in applying them to other Abelian theories containing usual vector
gauge fields coupled to matter fields in a
suitable way.\ Therefore, we intend here to apply the scheme discussed above
to a quite general Lagrangian which contains, among others, a
$\lambda\fy^4$-type vertex, in order to obtain its {\em generalized}, say,
point-split version.\ This new Lagrangian will be explicitly constructed up to
the second order in the gauge coupling constant. In addition, the present work
sets out not only to show the good applicability of this alternative
regularization procedure to a quite interesting class of Abelian interaction
vertex, but also to motivate further investigation towards its non-Abelian
version, which includes among others, the so-celebrated Higgs mechanism of the
Salam-Weinberg electroweak theory.\\ \\
Here, it is worthy noticing that the gauge transformation parameter may
explicitly appear in point-split actions, depending on the number of matter
fields involved in the interaction vertex. Actually, while for 3-vertices (two
matter plus one gauge field) the gauge parameter is generally absent from the
point-split action, in matter 4-vertices its presence turns out to be, as far
as we have understood, a natural ingredient to preserve gauge invariance under
generalized gauge transformations to a given order in the gauge
coupling constant. This comes from the fact that, in the framework of ggt's the
gauge invariance, in a generalized sense, has to be constructed and checked
order by order in a gauge coupling constant expansion. Such an issue will
become clearer throughout this work.\\ \\
Our paper follows the outline below. In Section 2 the Lagrangian which will
be worked out is presented as well as a survey of the point-splitting scheme
combined with generalized gauge transformations. Then, we apply such a
procedure to our Lagrangian and step-by-step we worked out its generalized
(split) expression up to the second order in the gauge coupling constant.
Section 3 is devoted to applications of the results obtained in the previous
section to some specific cases, say, $(\overline{\psi}\psi)^2$ and a modified
version of the Avdeev-Chizhov models. We close our paper
by pointing out some Concluding Remarks. Among others interests, we pay
attention to the applicability of our results to the Abelian Higgs model
whenever spontaneous symmetry breaking is concerned.  
\section{The Lagrangian and the regularization procedure} We shall start this
section by considering the following Lagrangian\ (which has the form of the
massive scalar Electrodynamics with self-interaction term, or the Abelian
Higgs model -provided that $m^{2}< 0$):\footnote{We shall use
Minkowski metric $diag(\eta_{\mu\nu})=(+,-,-,-)$ and greek letters running
0,1,2,3.}
\ba {\cal L}(x)=-\frac14 F_{\mu\nu}F^{\mu\nu}
+(D_{\mu}\fy)^{\dagger}(D^{\mu}\fy) -\frac{m^2}{2} \fyd\fy
-\frac{\lambda}{4}(\fyd\fy)^2 \label{1}\, , \ea  with $D_{\mu} =
\partial_{\mu} +ieA_{\mu}$ and $F_{\mu\nu}=\partial_{\mu} A_{\nu}
-\partial_{\nu} A_{\mu}$.\ Clearly, the matter fields are considered to be
complex\footnote{For further applications to fermionic fields, the Hermitian
conjugation must be changed to Dirac conjugation.\ On the other hand, if the
matter fields are rank-2 tensors, then additional attention must be paid to
their indices.\ See Section III for details. In addition, in dealing with the
actual Abelian Higgs model a important question that
now takes place is whether the present scheme is more suitable applied before
or after the spontaneous symmetry breaking be performed. Such a point will be
discussed later (see Concluding Remarks, for more details).} and their
products are taken at the same space-time point, say, $x$.\ This Lagrangian is
invariant under the usual local gauge transformations: 
\ba \delta
A_{\mu}(x)=-\partial_{\mu}\Lambda(x)\,; \qquad \delta\fy(x)=
+ie\Lambda(x)\fy(x) \, ; \qquad \delta\fyd(x)=-ie\Lambda(x)\fyd(x)\,
\label{2}. \ea
Now, in order to obtain a point-split version of the Lagrangian
above, i.e., a form free from same point product of fields in interaction
terms, we begin by writing the generalized version of the gauge
transformations, ggt's (denoted by $\delta_{g}$) up to $e^2$ (see
Ref.\cite{GNW} for further details): 
\ba & & \gt A_{\mu}(x)= -\partial_{\mu}
\Lambda(x)=\delta A_{\mu}(x) \label{gtA},\\ & & \gt\fy(x)=+ie \Lambda(1)\fy(2)
+\frac12 (ie)^{2} [\Lambda(1) +\Lambda(3)] (1,3)\fy(4) +{\cal O}(e^{3})\,
\label{gtfi},\\ & & \gt \fyd(x)= -ie \Lambda(-1)\fyd(-2) +\frac12 (ie)^{2}
[\Lambda(-1) +\Lambda(-3)] (-1,-3)\fyd(-4)+{\cal O}(e^{3})\, \label{gtfid},
\ea where we have defined:  \ba & & \Lambda(\pm n)=\Lambda(x\pm na);\qquad
\fy(\pm n)=\fy(x\pm na);\qquad \fyd(\pm n)=\fyd(x\pm na);\label{def1}\\ & &
(\pm m,\pm n)=\lim_{b\to 0^{+}} \int^{x\pm na-b}_{x\pm ma +b}
A_{\mu}(\eta)d\eta^{\mu} \, \label{def2}. \ea 
the point-splitting being
implemented by the (constant) 4-vector $a_\mu\equiv a$.\\ \\ 
From the definition above, we realize the first price to be paid in order to
avoid product of fields at the same point: the non-locality of the new
model. We shall come back to this point later.\\ \\
These ggt's can be shown to
satisfy the {\em generalized Abelian condition} up to $e^{2}$, i.e., the
commutator of two distinct ggt's (each of them with its respective parameter
$a_1$ and $a_2$) vanishes up to this order:
$$[\delta_{g1},\delta_{g2}]\fy(x)={\cal O}(e^{3}); \qquad
[\delta_{g1},\delta_{g2}]\fyd(x)={\cal O}(e^{3}).$$
is worthy noticing that, as the parameter $a$ is set to zero, all the results
above recover the usual ones (hereafter, by consistency, the same should
happen to all point-split results).\ Furthermore, we should stress that the
point-splitting acts only in transformations which present same point product,
which is the case for $\delta\fy$ and $\delta\fyd$, but not for $\delta
A_{\mu}$.\footnote{In the Abelian case, $\gt A_{\mu}=\delta A_{\mu}$ holds,
but in the non-Abelian scenario, where the ordinary gauge transformation for
$A^{a}_{\mu}$ involves products at the same point, the point-splitting will
act on it, and its non-Abelian ggt's will be different from the usual one.\
Indeed, such ggt's were already worked out for $SU(2)$ \cite{GNW2}, and more
recently for $SU(N)$ \cite{GWu}.}\\ \\ 
Now, we discuss the invariance of the ordinary Lagrangian, eq. (\ref{1}),
under the ggt's above (more precisely, up to order $e^{2}$).\ The kinetic
gauge term is clearly invariant since $\gt A_{\mu}=\delta A_{\mu}$.\ The mass
term for matter fields can be shown to be invariant in its action form, $\int
m^2 \fyd\fy d^{4}x$, with suitable change of variables within the integration
(see Ref. \cite{GNW} for more details).\ Contrary, the other terms are not
invariant and must have their points split up.\ We choose to do the
point-splitting (P.S) in the following way (like as in (\ref{def1}),
$A_{\mu}(\pm n)$ stands for $A_{\mu}(x \pm n)$):  \ba
&
(D_{\mu}\fy(x))^{\dagger}(D^{\mu}\fy(x)) \stackrel{P.S}{\longrightarrow}
&(D_{\mu}\fy)^{\dagger}(D^{\mu}\fy)_{P.S}=\\ \nn & & =\left[\partial_
{\mu}\fyd(x) -ie A_{\mu}(-1)\fyd(-2)\right]\left [\partial^{\mu}\fy(x) +ie
A_{\mu}(1)\fy(2)\right]\,, \label{8} \vspace{.5cm}\\ &
\left(\fyd(x)\fy(x)\right)^{2}\stackrel{P.S} {\longrightarrow}&
\left(\fyd\fy\right)^{2}_{P.S}= \fyd(-1)\fy(1)\fyd(-2)\fy(2) \, \label{psfi4}.
\ea And the split Lagrangian takes the form: \ba {\cal L}^{(0)}_{P.S}=
-\frac14 F_{\mu\nu}(x)F^{\mu\nu}(x) -\frac{m^2}{2} \fyd(x)\fy(x)
+(D_{\mu}\fy)^{\dagger}(D^{\mu}\fy)_{P.S}
-\frac{\lambda}{4}(\fyd\fy)^{2}_{P.S} \label{L0ps}\, . \ea
Here, it is worthy
noticing that, while the kinetic matter term,
$\partial_{\mu}\fyd(x)\partial^{\mu}\fy(x)$, involves a product at the same
point, it does not need to be split because the action of the ggt's on it will
produce regularized terms.\ Now, taking $\gt$ of such split terms up to order
$e$, we get: 
\ba \gt\left((D_{\mu}\fy)^{\dagger}(D^{\mu}\fy)_{P.S}\right)= (ie)\left[
\Lambda(1)\partial_{\mu}\fyd(x)\partial^{\mu}\fy(2) -\Lambda(-1)
\partial_{\mu}\fyd(-2)\partial^{\mu}\fy(x)\right] +{\cal O} (e^{2})\, . \nn
\ea
which is, at first glance, non-vanishing. But, if we take
its action form, we can perform a change of variables to show that the
integrals exactly cancel each other. In other words, the r.h.s. of the
previous expression gives rise to a vanishing term in the full split
action.\\Next, for the self-interaction term, we get:  \ba
& \gt\left((\fyd\fy)^{2}_{P.S}\right)= & ie\left[ \Lambda(2)\fyd(-1)\fy(3)
-\Lambda(-2)\fyd(-3)\fy(1)\right] \fyd(-2)\fy(2)\,+\\ \nn
& & +ie \fyd(-1)\fy(1)\left[ \Lambda(3)\fyd(-2)\fy(4)
-\Lambda(-3)\fyd(-4)\fy(2)\right] +{\cal O}(e^{2})\, \nn.
\ea
Contrary to the previous one, the term above seems to be intrinsically
non-vanishing; in fact, we did not see any way to set it to zero, neither by
a suitable change of variables nor by partial integration.\ Therefore, we must
search for a new term, $ \Omega^{(1)}_{P.S}$, such that $(\fyd\fy)^{2}_{P.S}+
\Omega^{(1)}_{P.S}$ be invariant under $\gt$ at least up to order $e$.\ This
term exists and can be explicitly written as: \ba
 \Omega^{(1)}_{P.S}=-ie \left( \{-2,2\}\fyd(-2)\fy(2)
+\{-3,3\}\fyd(-1)\fy(1)\right) \label{omega1}\, ,
\ea
with the definition: 
\ba
\{-n,+n\}= \lim_{b\to 0^+}\int^{x+na-b}_{x-na+b}dy^{\mu} \partial_{\mu}\left[ 
\fyd(\frac{y}{n} +\frac{n-1}{n}x -na)\,\fy(\frac{y}{n}
+\frac{n-1}{n}x +na)\,(-\infty,y)\right]\,, \label{chaven}
\ea
where $(-\infty,y)$ stands for $\int^{y}_{-\infty}\,
A^{\nu}(\eta)d\eta_{\nu}$.\\Therefore, the split Lagrangian, whose action is
invariant under $\gt$ up to first order, ${\cal L}^{(1)}_{P.S}$, is the sum of
${\cal L}^{(0)}_{P.S}$ and $ -\frac{\lambda}{4}\Omega^{(1)}_{P.S}$
(eqs. (\ref{L0ps}) and (\ref{omega1})).\\ \\
It is precisely in this sense that gauge invariance has to be taken in the framework
of generalized gauge transformations, ggt's. Actually, since the ggt's themselves
take the form of an infinite series in the gauge coupling constant, then it is
expected that the split (and regularized) action also presents a
similar form, with its ``{\em generalized gauge invariance} being constructed
and checked order by order.\\ \\
Now, calculating $\gt{\cal L}^{(1)}_{P.S}$ at order $e^2$, we get (after suitable
change of variables in the action form of the terms):
\ba
\gt\left((D_{\mu}\fy)^{\dagger}(D^{\mu}\fy)_{P.S}\right)\arrowvert_{e^2}=
(ie)^{2}\left[ \Lambda(1)A_{\mu}(3)-\Lambda(3)A_{\mu}(1)\right]\fyd(x)\stackrel
{\leftrightarrow}{\partial^{\mu}} \fy(4) \, ,
\ea
(with $U\stackrel{\leftrightarrow}{\partial} V = U\partial
V-(\partial V)U$).\ Again, we cannot set this term to zero.\ Instead, according to
$\Omega^{(1)}_{P.S}$, we must search for a new term, $ \Sigma^{(2)}_{P.S}$, such
that $(D_{\mu}\fy)^{\dagger}(D^{\mu}\fy)_{P.S} + \Sigma^{(2)}_{P.S}$ be invariant
under $\gt$ at least up to order $e^2$.\ Such term can be found and its simplest
form is: $$ \Sigma^{(2)}_{P.S}=-(ie)^2 \,\Sigma_{\mu}\, \left[\fyd(x)
\stackrel{\leftrightarrow}{\partial^{\mu}} \fy(4)\right],$$ with
$\Sigma_{\mu}$ being a function of $\Lambda$ and $A_{\mu}$.\ In fact,
$\Sigma_{\mu}$ must be an object such that $\gt\Sigma_{\mu}= \Lambda(1)
A_{\mu}(3)-\Lambda(3)A_{\mu}(1)$.\ It is easy to check that the
following expression satisfies such a requiriment:
\ba
 \Sigma^{(2)}_{P.S}=-(ie)^2\left\{[A_\mu (1)+A_\mu(3)]\,(1,3) \,+([1]+[3])
\int_{x+a}^{x+3a} F_{\mu\nu}(\xi)d\xi^\nu\right\}\fyd(x)\stackrel{\leftrightarrow}
{\partial^{\mu}} \fy(4) \label{sigma2} \, ,
\ea
with $[\pm n]= \frac12 \left[ (-\infty,\pm n) + (\infty,\pm n)\right]$.\ In
addition, we may see that as $a\to 0$ then $ \Sigma^{(2)}_{P.S}$
vanishes.\ [The expression inside $\{\,\}$ was already obtained in Ref.\cite{GNW}
for the interacting vertex of QED; for details, see eq. (24) in that paper]\footnote{An
alternative, but apparently non-equivalent, form for $\Sigma^{(2)}_{P.S}$
was obtained in Ref.\cite{cax96} and reads:$$-(ie)^{2}\left\{[1]A_{\mu}(3)
-[3]A_{\mu}(1)
-\frac12 \left( \frac{[1]^2}{\Lambda(1)}\partial_{\mu}\Lambda(3) -
\frac{[3]^2}{\Lambda(3)}\partial_{\mu}\Lambda(1) \right)\right\}
\fyd(x)\stackrel{\leftrightarrow}{\partial^{\mu}} \fy(4).$$Despite the
explicit presence of the gauge parameter in the expression above, it can be
verified that it vanishes as $a\to0$ and leads us to a split action invariant
 under ggt's up to the 2nd order.}.\\ \\
Now, for the self-interaction sector, we get:
\ba
& \gt& \hspace{-.2cm}\left((\fyd\fy)^{2}_{P.S}+ \Omega^{(1)}_{P.S}\right)
\arrowvert_{e^2}= \frac12 (ie)^{2}\left\{\,2\,\{-2,2\}\,\left[
\Lambda(-3)\fyd(-4)\fy(2) -\Lambda(3)\fyd(-2)\fy(4)\right] \, \right.+\nn \\
& & \left.+\,2\,\{-3,3\}\,\left[ \Lambda(-2)\fyd(-3)\fy(1)
-\Lambda(2)\fyd(-1)\fy(3)\right]\, +\right. \nn \\
& &\left.-[\Lambda(-2)+\Lambda(-4)] \,(-2,-4)\left[\fyd(-5)\fy(1)\fyd(-2)\fy(2)
+\fyd(x)\fy(2)\fyd(-5)\fy(3)\right]+\right. \nn \\ 
& & \left. +\,[\Lambda(2)+\Lambda(4)]\,(2,4)\left[\fyd(-1) \fy(5)\fyd(-2)\fy(2)
+\fyd(-2)\fy(x)\fyd(-3)\fy(5)\right]\right.+\nn \\
& & \left. -\,2\,\Lambda(4)\, (-\infty,2)\left[\fyd(-1)\fy(5)\fyd(-2)\fy(2)
+\fyd(-3)\fy(5)\fyd(-2)\fy(x)\right]\right.+\nn \\
& & \left.  -\,2\,\Lambda(-4)\, (-\infty,-2)\left[\fyd(-5)\fy(1)\fyd(-2)\fy(2)
+\fyd(-5)\fy(3)\fyd(x)\fy(2)\right] \right.+\nn \\
& & \left. +\,2\,\left[ \Lambda(2)\,(-\infty,-2) +\Lambda(-2)\,(-\infty,2)
\right]\fyd(-3)\fy(3)\fyd(-2)\fy(2)\, +\right.\nn \\
& & \left. +\,2\,\left[ \Lambda(3)\,(-\infty,-3) +\Lambda(-3)\,(-\infty,3)
\right]\fyd(-4)\fy(4)\fyd(-1)\fy(1)\,\right\}\,.\label{eq16}
\ea 
The non-vanishing of this term is evident.\ Searching for a new term,
$ \Omega^{(2)}_{P.S}$, such that $(\fyd\fy)^{2}_{P.S} + \Omega^{(1)}_{P.S}+
\Omega^{(2)}_{P.S}$ be invariant under $\gt$ at least up to order $e^2$, is more
difficult than for the former ones, $ \Omega^{(1)}_{P.S}$ and
$\Sigma^{(2)}_{P.S}$.\ Such difficulties arise from its rather complicated
structure.\ Fortunately,  an explicit expression may be indeed found.\ For that,
we notice that the six last terms above have similar structure, say,
$\Lambda(\pm n) \, (\pm m,\pm p) \,\fyd\fy\fyd\fy$-type factors.\ Actually, for
such terms, the simplest $ \Omega^{(2)}_{P.S}$-type `{\em counter-term}' has the general
form:$$\frac12 (ie)^{2}\left(\frac12 \frac{\Lambda(\pm n)}{\Lambda(\pm p)
-\Lambda(\pm m)}\, (\pm m,\pm p)^{2}\right).\\$$
By remembering the definitions of the above quantities, it is easy to see that such
a expression vanishes as $a\to 0$.\\
On the other hand, for the first two terms in eq. (\ref{eq16}), those proportional
to $\{-n,+n\}$, the task of finding $ \Omega^{(2)}_{P.S}$-type {\em counter-terms} appear
to be very easy if we take into account that:$$\gt\{-n,+n\}\, \arrowvert_{e^0}= 
\Lambda(-n)\fyd(-n-1)\fy(n-1) -\Lambda(n)\fyd(-n+1)\fy(n+1).$$In fact, as it can be
readily checked, those first two terms have the following $ \Omega^{(2)}_{P.S}$-type
{\em counter-term}:\footnote{In fact, if $\fy$ (and $\fyd$) are fermionic
fields, then $\fy$ (or $\fyd$) has anticommutative
property, but the bilinear $\fyd\fy$ has commutative behavior.\ Therefore, even for
fermionic fields, we can change the order of $\{-2,+2\}$ by $\{-3,+3\}$ and
vice-versa, without any extra minus sign.}$$\frac12 (ie)^2 \left( -2 \,\{-2,+2\}\,
\{-3,+3\}\right).\\$$
Therefore, the full $ \Omega^{(2)}_{P.S}$-term takes over the form:
\ba
& \Omega^{(2)}_{P.S}= &-\frac12 (ie)^2 \left[ \quad 2\,\{-2, +2\}\, \{-3,+3\}
\quad+\right.\nn \\
& & \left.+\left( \frac{\Lambda(2)+\Lambda(4)} {\Lambda(2)-\Lambda(4)}\right)
\frac{(2,4)^2}{2} \left[\fyd(-1)\fy(5)\fyd(-2)\fy(2) +\fyd(-2)\fy(x)\fyd(-3)
\fy(5)\right] +\right.\nn \\
& & \left.\hspace{-.2cm}-\left(\frac{\Lambda(-2)+\Lambda (-4)}{\Lambda(-2)
-\Lambda(-4)}\right) \frac{(-2,-4)^2}{2} \left[\fyd(-5)\fy(1)\fyd(-2)\fy(2)
+\fyd(x)\fy(2)\fyd(-5)\fy(3)\right]+\right.\nn \\
& & \left. +\,\frac{\Lambda(4)}{\Lambda(2)}\, (-\infty, 2)^2 \left[\fyd(-1)
\fy(5)\fyd(-2)\fy(2) +\fyd(-3)\fy(5)\fyd(-2)\fy(x)\right]+\right.\nn \\
& & \left. +\,\frac{\Lambda(-4)}{\Lambda(-2)}\, (-\infty, -2)^2
\left[\fyd(-5)\fy(1)\fyd(-2)\fy(2) +\fyd(-5)\fy(3)\fyd(x)\fy(2)
\right]+\right.\nn \\
& & \left. -\,\left( \frac{\Lambda(2)}{\Lambda(-2)}\, (-\infty, -2)^2
+\frac{\Lambda(-2)}{\Lambda(2)}\, (-\infty, 2)^2 \right)\fyd(-3)\fy(3)
\fyd(-2)\fy(2)+\right.\nn \\
& & \left. -\, \left( \frac{\Lambda(3)}{\Lambda(-3)}\, (-\infty, -3)^2
+\frac{\Lambda(-3)}{\Lambda(3)}\, (-\infty, 3)^2 \right)\fyd(-4)\fy(4)
\fyd(-1)\fy(1) \,\right]\label{omega2}\,.
\ea
Finally, the ${\cal L}^{(2)}_{P.S}$ Lagrangian, whose action is invariant under
$\gt$ up to order $e^2$, may be written as:
\ba
{\cal L}^{(2)}_{P.S}={\cal L}^{(0)}_{P.S}+ \Sigma^{(2)}_{P.S} -\frac{\lambda}{4}
(\Omega^{(1)}_{P.S}+ \Omega^{(2)}_{P.S}) \label{L2ps}\,,
\ea
with the expressions for the above terms being given by (\ref{L0ps}),
(\ref{omega1}), (\ref{sigma2}), and (\ref{omega2}).\\ \\
The form of eq. (\ref{L2ps}) deserves further remarks. The explicit presence
of the gauge parameter, $\Lambda$, in eq. (\ref{omega2}), may seem to be
spurious, since it is well-known that (usual) gauge invariance is explicitly
broken by the presence of the gauge parameter in the action. Actually, such
a symmetry is broken whenever it is taken in the usual sense, but by reassessing
the meaning of gauge invariance in the
context of ggt's, then the scenario may be changed. As we have already pointed
out, gauge invariance has in this framework to be constructed and checked
by means of an order by order (in the gauge coupling constant) algorithm.\\ \\
Therefore, when finding out a
split action, this has to be also done order by order and its `{\em generalized
gauge invariance}' must be verified according such a ``perturbative''
procedure. Hence, if we wish to verify whether a given split action is
invariant under ggt's, say, up to 2nd order, for concreteness, then we
must check: first, if as the splitting parameter vanishes, $a\to0$,
the split action restores the original one; and, if the split action is
actually invariant under ggt's up to 2nd order, then
$\gt S^{(2)}_{P.S.}={\cal O}(e^3)$.\\ \\
In our present case, eq.(\ref{L2ps}), both requirements are verified, even
though the gauge parameter is explicitly present in its expression.
Whether other good split
actions without explicit presence of such a parameter may be found for
$\lambda\fy^4$-type vertices is not so clear to us. Indeed, in the case
of 3-leg vertices, like as $\fyd A_\mu\gamma^\mu\fy$, we have found a
$\Lambda$-explicitly dependent action which was shown to satisfy both
of the requiriments above (see footnote that follows eq.(\ref{sigma2})).\\   
\section{Applications to some self-interacting models}
Here, in order to illustrate the applicablity of our results, we shall deal
with some $\lambda\varphi^4$-type models.\ Whenever necessary, we shall pay
attention to specific points which were not still presented.\\ \\
i) {\bf The $(\overline{\psi}\psi)^2$ model}\\ The model which will be worked
out is described by the following Lagrangian:\footnote{It is worthy noticing
the (power-counting) non-renormalizability of this self-interaction vertex:
$[g]=[mass]^{-1}$ in (3+1) dimensions.} \ba {\cal L}_{\psi}(x)=-\frac14 F^{\mu\nu}F_{\mu\nu}
+\overline{\psi}\left(  iD_{\mu}\gamma^{\mu} -m_f\right)\psi
-g(\overline{\psi}\psi)^2 \label{Lpsi}\, , \ea with $D_{\mu}$ and $F_{\mu\nu}$
previously defined.\\ \\
Here, due to the anti-commutative character of
fermionic fields, we must pay special attention in changing the order of such
fields.\ Moreover, the kinetic term is slightly different from that for scalar
field and it must be taken apart.\ Fortunately, such a term was already
studied in Ref.\cite{GNW} and, if we perform the following splitting: \ba
&\overline{\psi}(x)\,iD_{\mu}\gamma^{\mu}\psi(x)\stackrel{P.S}
{\longrightarrow}\, &(\overline{\psi}\,iD_{\mu}\gamma^{\mu}\psi)_{P.S}= \nn \\
& & \overline{\psi}(x)\,i\partial_{\mu}\gamma^{\mu}\psi(x)
-e\overline{\psi}(x-a) A_{\mu}(x)\gamma^{\mu}\psi(x+a)\nn \, ,
\ea
one can readily show that $\int {d^{4}x}\,(\overline{\psi}\,iD_{\mu}
\gamma^{\mu}\psi)_{P.S}$ is invariant up to order $e$.\ At second order, such a
variation does not vanish, but it is exactly canceled by the following term
(quite similar to eq.(\ref{sigma2}) above; see also eq. (24) in Ref.\cite{GNW}):
\ba
\Sigma^{(2)}_{\psi,\,P.S}= -\frac{ie^2}{2} \overline{\psi}(-2) \gamma^{\mu}
\psi(2) \, \left\{[A_{\mu}(-1)+A_{\mu}(1)](-1,+1)+([-1]+[1])\int^{x+a}_{x-a}d
\eta^{\nu}F_{\mu\nu}(\eta)\right\} \label{sigma2psi}\, .
\ea
Now, the $(\overline{\psi}\psi)^2$-term is split in the same way as
$(\fyd\fy)^2$: $$(\overline{\psi}(x)\psi(x))^2\stackrel{P.S}{\longrightarrow}
\,(\overline{\psi}\psi)^2_{P.S}= \overline{\psi}(-1)\psi(1)\overline{\psi}
(-2)\psi(2).$$\\
So, ${\cal L}^{(0)}_{P.S}$ for $\psi$-like fields reads:
\ba
{\cal L}^{(0)}_{\psi,\,P.S}=-\frac14 F^{\mu\nu}(x)F_{\mu\nu}(x) -m_f
\overline{\psi}(x)\psi(x) +i(\overline{\psi}D_{\mu}\gamma^{\mu}\psi)_{P.S}
-g(\overline{\psi}\psi)^2_{P.S} \label{L0psi}\, .
\ea
To get ${\cal L}^{(2)}_{\psi,\,P.S}$, we may use the $\Omega^{(1)}_{P.S}$
and $\Omega^{(2)}_{P.S}$ obtained in the previous section with suitable
change of $\fy$ by $\psi$ and $\fyd$ by $\overline{\psi}$.\ Indeed, as we kept
the original order of those matter fields in the previous results, we may write:
\ba
\Omega^{(1)}_{\psi,\,P.S}=\Omega^{(1)}_{P.S}\arrowvert_{\fy\to\psi,\,\fyd
\to\overline{\psi}}\quad \mbox{and}\quad \Omega^{(2)}_{\psi,\,P.S}=
\Omega^{(2)}_{P.S}\arrowvert_{\fy\to\psi,\,\fyd
\to\overline{\psi}}\label{omegaspsi}\, .
\ea
Finally, we get:
\ba
{\cal L}^{(2)}_{\psi,\,P.S}={\cal L}^{(0)}_{\psi,\,P.S} +\Sigma^{(2)}_{\psi,\,P.S}
+g(\Omega^{(1)}_{\psi,\,P.S}+\Omega^{(2)}
_{\psi,\,P.S})\label{L2psi}\,.\\\nn
\ea 
ii){\bf The `Avdeev-Chizhov' model}\\ Some years ago, Avdeev and
Chizhov\cite{AC} proposed an Abelian model which includes antisymmetric rank-2
real tensors that describe matter, rather than gauge degrees of freedom.\ They
are coupled to a usual vector gauge field as well as to fermions.\ The model
was shown to reveal interesting properties: for instance, these new matter
fields were shown to play an important role in connection with extended
electroweak models in order to explain some observed decays like
$\pi^-\to e^- +\overline{\nu}+\gamma$ and $K^+\to \pi^0 +e^+ +\nu$
\cite{Chi}, and a classical analysis of its dynamics has shown that some
longitudinal excitations may carry ``physical degrees of freedom ''
(see Ref.\cite{AC2}, for further details).\ In addition, some works have been
devoted to the study of its supersymmetric generalization \cite{NogCia}, as
well as its connection with non-linear sigma models \cite{NPFH}.\\ \\
Starting off from these interesting features, it was shown that the coupling
between tensorial and fermionic fields generates anomalies in the quantized
version of the model and could also spoil its renormalizability \cite{LRS}.\
The removal of the fermions has the additional usefulness of allowing us to
write the new Lagrangian in a shorter form by means of complex field tensors,
$\fy_{\mu\nu}$ and $\fyd_{\mu\nu}$ \cite{LRS2}.\ Thus, the {\em modified}
Avdeev-Chizhov model reads:
\ba
{\cal L}_{AC}(x)= -\frac14 F_{\mu\nu}F^{\mu\nu} +(D_{\mu}\fy^{\mu\nu})
(D^{\alpha}\fy_{\alpha\nu})^{\dagger} -\frac{\lambda}{4}\fyd_{\mu\nu}
\fy^{\nu\kappa}\fyd_{\kappa\lambda}\fy^{\lambda\mu}
\label{Lac}\,, 
\ea
with $D_\mu$ and $F_{\mu\nu}$ already defined.\ Once $\fy_{\mu\nu}$ is taken to
satisfy a complex self-dual relation: $$\fy_{\mu\nu}(x)=+i\tilde{\fy}_{\mu\nu}(x)
\, , \qquad \tilde{\fy}_{\mu\nu}=\frac12 \epsilon_{\mu\nu\alpha\beta}\fy^
{\alpha\beta},$$then it can be split into two (real tensors) parts:$$\fy_{\mu\nu}
(x)=T_{\mu\nu}(x) +i \tilde{T}_{\mu\nu}(x)\quad\mbox{and}\quad \fyd_{\mu\nu}(x)=
T_{\mu\nu}(x) -i \tilde{T}_{\mu\nu}(x),$$where $T_{\mu\nu}$ and $\tilde{T}_{\mu\nu}$
are real and antisymmetric fields.\ Actually, regarded as matter fields, it is
known that they describe spin$\_$0 excitations (see, for example, Refs.
\cite{AC2,nogue}).\\ \\ 
Now, performing similar splittings in ${\cal L}_{AC}(x)$, as we have done in
former cases, we get, after some calculation, ${\cal
L}^{(2)}_{P.S}(\fy^{\mu\nu})$: \footnote{We have already found a similar
expression for this model in Ref.\cite{cax96}.\ There, a slightly modified
splitting was employed, as well as a lengthier form for
$\Sigma^{(2)}_{P.S}(\fy^{\mu\nu})$.}  
\ba
{\cal L}^{(2)}_{P.S}(\fy^{\mu\nu})={\cal L}^{(0)}_{P.S}(\fy^{\mu\nu})
+\Sigma^{(2)}_{P.S}(\fy^{\mu\nu}) -\frac{\lambda}{4} \left(\Omega^{(1)}_{P.S}
(\fy^{\mu\nu})+ \Omega^{(2)}_{P.S}(\fy^{\mu\nu})\right)
\ea
where the terms above have the following expressions:
\ba
{\cal L}^{(0)}_{P.S}(\fy^{\mu\nu})=-\frac14 F_{\mu\nu}F^{\mu\nu}
+(D_{\mu}\fy^{\mu\nu})_{P.S}(D^{\alpha}\fy_{\alpha\nu})^{\dagger}_{P.S}
-\frac{\lambda}{4}(\fyd_{\mu\nu}\fy^{\nu\kappa}\fyd_{\kappa\lambda}
\fy^{\lambda\mu})_{P.S}
\ea
(with the splittings previously performed);
\ba
&\Sigma^{(2)}_{P.S}(\fy^{\mu\nu})=-(ie)^2 &\left[ [1]A^{\mu}(3)
-[3]]A^{\mu}(1)-\frac12\left(\frac{[1]^2}{\Lambda(1)}\partial^{\mu}
\Lambda(3)-\frac{[3]^2}{\Lambda(3)}\partial^{\mu}\Lambda(1)\right)
\right]{\mbox{\small{\rm x}}} \nn\\
& & {\mbox{\small{\rm x}}}\left[\fyd_{\mu\nu}(x)\partial_{\alpha}\fy^{\alpha\nu}(4)
-\fy_{\mu\nu}(4)\partial_{\alpha}\varphi^{\dagger\alpha\nu}(x)\right]
\ea
(its slight difference with respect to $\Sigma^{(2)}_{P.S}$,
eq. (\ref{sigma2}),
is due to the tensor indices);
\ba
\Omega^{(1)}_{P.S}(\fy^{\mu\nu})=(-ie)\left( \{-2,+2\}^{\mu\hspace{.07in}}_\nu
\fyd_{\mu\alpha}(-2)\fy^{\alpha\nu}(2) +\{-3,+3\}^{\mu\hspace{.07in}}_\nu
\fyd_{\mu\alpha}(-1)\fy^{\alpha\nu}(1)\right)\,,
\ea
and $\Omega^{(2)}_{P.S}(\fy^{\mu\nu})$ which is easily obtained from
$\Omega^{(2)}_{P.S}$ by making the interchanges:
\ba
& & \{-2,+2\}\{-3,+3\}\longrightarrow \{-2,+2\}^{\mu}\mbox{}_\nu
\{-3,+3\}^{\nu}\mbox{}_\mu \nn\\
& & \fyd\fy\fyd\fy\longrightarrow \fyd_{\mu\nu}\fy^{\nu\kappa}\fyd_{\kappa\alpha}
\fy^{\alpha\mu}\,, \nn
\ea
where we have defined $\{-n,+n\}_\mu\mbox{}^\nu$ in the same way as $\{-n,+n\}$,
with $\fyd\fy$ changed by $\fyd_{\mu\alpha}\fy^{\alpha\nu}$ in its definition,
eq. (\ref{chaven}).\\ \\
Usually, the 2-form field, $\varphi_{\mu\nu}$, is treated as a gauge potential
(the so-called Cremmer-Scherk-Kalb-Ramond field \cite{CS,KR}), and in (3+1)
dimensions it describes a massless scalar excitation.\ However, when mixed to
the Maxwell field, $A_\mu$, by means of a topological mass term that linkes two
Abelian factors, it yields a massive spin$\_$1 excitation in the spectrum.\ On
the other hand, we should stress that its coupling to charged matter may be
realized only non-minimally, i.e., by means of its (3-form) field-strength
\cite{nossos}.\\ \\
Furthermore, from the point of view of the point-splitting procedure, since its
usual gauge transformation, $\delta\varphi_{\mu\nu}(x)=\partial_\mu\xi_\nu(x)
-\partial_\nu\xi_\mu(x),$ does not involve products of quantities at the same
space-time point, such a tranformation does not undergo any change in going to
the generalized case, say, $\delta_g\varphi_{\mu\nu}(x)=\delta\varphi
_{\mu\nu}(x)$.\ Therefore, by viewing $\varphi_{\mu\nu}$ as a gauge potential,
its ggt's take easier expressions than when it is treated as a matter field
(a general fact, at least in the Abelian framework, in dealing with such a
procedure).\ This readily implies remarkable simplifications whenever working
with $\varphi_{\mu\nu}$ as a gauge potential.\\
\section{Concluding Remarks} The point-splitting procedure
combined with generalized gauge transformations has yielded regularized
Lagrangians which contains $\lambda\fy^4$-type interaction.\ The result is such
that the generalized Lagrangians have their interacting terms defined at
different space-time points.\ Nevertheless, this property introduces non-locality
at the level of the regularized theory.\\ \\
In general, non-local theories cannot be quantized with the usual methods and the
interpretation of their results are not quite obvious.\ Moreover, we know that
non-locality can lead to troubles as long as the causality of the theory is
concerned. However, these problems arise only for the regularized theory, in
much the same way as ghosts are present and unitarity is temporarily lost for
regularized theories before the regularization parameter is removed.\\ \\
Nevertheless, Osland and Wu \cite{OW} obtained some standard results in QED
starting by a split Lagrangian (with regularity and non-locality
properties)\footnote{It is worthy noticing that their Lagrangian (eq. (2.7)
in Ref.\cite{OW}) is different from the `{\em correct}' generalized
QED-Lagrangian, up to fourth order (eq. (24) in Ref.\cite{GNW}).\ Such a
difference may be explained by noticing that, in Ref. \cite{OW}, the generalized
gauge covariance is not taken in its precise meaning.}.\ Their method works for
the calculation of the quantities with a dependence in the splitting parameter,
which is set to zero at the end of calculations in order to get the standard
results.\\ \\
What we may learn from these calculations is that, when point-splitting is
combined with generalized gauge transformations in order to obtain regularized
(Abelian) Lagrangians, the task becomes more difficult with the increasing of
the number of matter fields at the same vertex; in general, the complications
which arise from the presence of extra (Abelian) gauge fields are minor
ones.\ So, the calculations involving $\lambda\fy^4$-type vertices are harder to
be performed than for `lower vertices', $\fy A_\mu\fy$,
$\fy A_{\mu}A^{\mu}\fy$, and so forth.\ In addition, higher-order terms in the
coupling constant are, in general, more complicated to be handled than lower
ones.\\ \\
Our present study is a good example of such complications and the reason why
they arise. For instance, in dealing with the 2nd order calculations of the
split version of the self-interacting vertex we have seen that a {\em
counter-term} for that term involved the explicit presence of
the gauge parameter, scenario which is expected to become even more intricate
at higher orders.\\ \\ 
Another point that should be stressed is that this procedure is independent
of the space-time dimension, and so, of the canonical dimension of the fields
(matter or gauge ones)\footnote{In fact, the ggt's depend on the splitting
parameter, the constant vector $a_\mu$, and on the Abelian (or non-Abelian)
character of the gauge fields.}.\ Hence, the expressions for our $\Sigma$ and
$\Omega$ terms remain valid in other dimensions. Therefore, our present
results could be equally well applied to four-matter (scalar, spinorial, and
rank-2 tensorial) vertices in lower or higher space-time dimensions.
However, special attention should be paid if dimensional
reduction and/or compactification, spontaneous symmetry breaking, or other
mechanisms are involved. As we shall discuss below for the case of the Higgs
model, the present procedure is suitably applied only at the stage in which
the true physical excitations are taken into account.\\ \\   
On the other hand, if we are dealing with a renormalizable theory (scalar, for
simplicity) in $(2+1)$ dimensions, an extra $f\fy^6$-term is allowed.\ In this
case, our results could be applied to the model, including the
$\lambda\fy^4$-term, but the extra term should be worked out apart. As we have
already said, the task of working out the split version of a matter vertex
tends to become more difficult as the number of matter fields increases. Thus,
we expect even more work in dealing with $\fy^6$-like matter vertices than we
had in the present case. Still concerning possible relevant applications in
this space-time, we may think of applying the present procedure for studying
some Abelian (and non-Abelian in a further stage, too)  models connected with
the Chern-Simons term. For instance, we may study some points concerning the
radiation produced by accelerated point-like charges\cite{inprogress}, an
issue which still demands several answers (see Ref.\cite{teseprd}, for more
details). \\ \\
Another relevant question that we may rise up here is the issue of the
point-splitting in connection with the Abelian Higgs mechanism.\ The
spontaneous symmetry breaking, as realized by a charged scalar, obliges a
shift of the Higgs field around its vacuum expectation value and induces the
appearance, among others, of trilinear matter couplings not present in the
original action.\ Here, we may wonder whether the splitting should be
performed before or after the breaking.\ Indeed, although our results are
directly applied to the unbroken phase, it does not seem to be the best
choice.\ Actually, we claim that the most suitable way to implement
point-splitting is after the breaking takes place, for in the broken regime
all possible vertices show up and we can really control the theory we are
dealing with, since only at this stage we are quantizing the truly physical
excitations.\ In this case, while our results are applicable to some terms of
the Lagrangian written around the true ground state, such as the kinetic and
quartic ones, the trilinear vertex, in turn, should be worked out apart
(expected to be of easier manipulation than the present one).\\ \\
We also hope that the present paper could help us whenever dealing with the
non-Abelian case. In this scenario novel features will arise mainly because
$\gt A^{a}_{\mu}$ will take more complicated (and lengthier) forms, and they
will imply in new ggt's for the matter fields which, in turn, will also take
lengthier expressions than those for the Abelian case.\\ \\ 
Furthermore, in view of the special role that supersymmetry and supersymmetric
gauge theories play in the programme of building up fundamental interaction
models, it  would be advisable to extend the point-splitting method to treat
supersymmetric theories in superspace. Point-split super-actions both in the
space-time and Grassmann coordinates may be an interesting issue since now
gauge invariance and supersymmetry must be simultaneously checked and many
features of the method must be revealed: the advantage of implementing the
point- -splitting procedure in superspace is that supersymmetry is manifest
and one needs not checking Ward identities (as it would be the case in a
component-field approach) to undertake that supersymmetry is kept upon
point-splitting.\\ \\ 
we claim that some questions concerning this issue should eventually become
clearer.\ For example, how could Feynman rules for such a kind of Lagrangian
be formulated? Or still, as we may see, there are some new ` interaction
terms' in the generalized Lagrangian.\ Could these new terms have some
physical interpretation and/or relevance?\\ \\ 
\centerline{\bf Acknowledgments}\vspace{.2cm} The authors are grateful to Dr.
A.L.M.A. Nogueira for useful discussions concerning the Avdeev-Chizhov model.
WAMM is grateful to CNPq and FAPEMIG for the financial support. JAHN thanks
CNPq for partial financial support.\\
 \end{document}